\documentclass[12pt, letterpaper, notitlepage]{iopart}
\usepackage{graphicx,color}

\begin{document}
\newcommand{\kk}{{\bf k}}
\newcommand{\Q}{{\bf Q}}
\newcommand{\q}{{\bf q}}
\newcommand{\gk}{g_\textbf{k}}
\newcommand{\ee}{\tilde{\epsilon}^{(1)}_\textbf{k}}
\newcommand{\HH}{\mathcal{H}}
\newcommand{\oned}{quasi-1D }
\newcommand{\twod}{2D }

\title[Spin-Orbit Coupling in LaAlO$_{3}$/SrTiO$_{3}$]{Spin-Orbit Coupling in LaAlO$_3$/SrTiO$_3$ interfaces: Magnetism and Orbital Ordering}
\author{Mark H. Fischer$^1$, Srinivas Raghu$^2$, Eun-Ah Kim$^1$}
\address{%
$^1$ Department of Physics, Cornell University, Ithaca, New York 14853, USA
}
\address{%
$^2$ Department of Physics, Stanford University, Stanford, California 94305, USA
}
\ead{mark.fischer@cornell.edu}

\date{\today}

\begin{abstract}
 Rashba spin-orbit coupling together with electron correlations in
  the metallic interface between SrTiO$_3$ and LaAlO$_3$ can lead to an unusual combination of magnetic and orbital ordering.
We consider such phenomena in the context of the recent observation of anisotropic magnetism.
Firstly, we show that Rashba spin-orbit coupling can account for the observed magnetic anisotropy, assuming a correlation driven (Stoner type) instability toward ferromagnetism. 
Secondly, we investigate nematicity in the form of an orbital imbalance between d$_{xz}$ / d$_{yz}$ orbitals. 
We find an enhanced susceptibility toward nematicity due to the van Hove singularity in the low-electron-density regime.
In addition, the coupling between in-plane magnetisation anisotropy and nematic order provides an effective symmetry breaking field in the magnetic phase. We estimate this coupling to be substantial in the low-electron-density regime.
The resulting orbital ordering can affect magneto transport.

\end{abstract}

\maketitle

\section{Introduction} 
Rashba spin-orbit-coupling (SOC) effects have been mostly studied in weakly interacting systems such as semiconductor heterostructures designed for spintronics applications\cite{zutic:2004}, and occur  in two-dimensional systems without mirror symmetry\cite{winkler:2003}. 
However, 
the effects of Rashba SOC 
in a two-dimensional system with strongly interacting electrons, found for example in interfaces, are emerging as a new frontier. 
There is thus a pursuit for new emergent phases of matter in this regime, both theoretically\cite{berg:2012} and experimentally\cite{hwang:2012}, with
the electron gas 
at the interface between the two non-magnetic insulators LaAlO$_3$ and SrTiO$_3$ (LAO/STO) the widest-studied such example. The observed ferromagnetic instability at this interface\cite{brinkman:2007} could well be of Stoner type, which suggests that electronic correlations may be enhanced due to low dimensionality and poor screening at low densities. Hence, combined with its demonstrated tunability\cite{caviglia:2008,reyren:2007}, the LAO/STO interface is an ideal testbed for physics of Rashba SOC in correlated electron systems. 


Recent observation of magnetic anisotropy may signal further richness in the phase diagram of the interface.  
Specifically, Bert \textit{et al.}~\cite{bert:2011} and Li \textit{et al.}~\cite{li:2011} observed strong in-plane preference for magnetisation [see figure \ref{fig:disp} (a)]. Bert \textit{et al.} attributed this observed anisotropy to the shape anisotropy of the interface: 
energetic bias towards a certain magnetisation direction e.g., along the longest axis of an ellipsoid,
driven by 
an anisotropic demagnetisation field.
However, for ultra thin films consisting of only a few atomic layers this effect is subdominant next to 
microscopic effects\cite{bland:2005,draaisma:1988}.

If the shape anisotropy is negligible in the interfaces, the dominant source of magnetic anisotropy would be spin-orbit coupling.
However, the typical alignment of the spin  with the largest angular momentum
in itinerant d-electron systems\cite{laan:1998} would predict an out-of-plane magnetisation upon occupation of d$_{xz}$ and  d$_{yz}$ orbitals, in disagreement with experiments\cite{joshua:2011,li:2011,bert:2011}.
In this work, we show that Rashba spin-orbit coupling leads to the unusual circumstance of an
 {\it anisotropic (spin) susceptibility}.
Assuming Stoner ferromagnetism close to a 
van Hove singularity near band edges due to SOC, we argue that this in turn leads to a magnetisation anisotropy.

We further investigate the possibility of nematic order in the form of orbital ordering between the d$_{xz}$ and  d$_{yz}$ orbitals, since the large density of states near the band edge also leads to an enhanced tendency towards such ordering.
This order can couple to an in-plane magnetisation anisotropy and we show here that a consequence of the Rashba SOC is a strong such coupling in the low-density regime near the band edge. This implies that orbital ordering accompanies the magnetic phase and leads to an additional magnetisation anisotropy within the plane.


\section{Anisotropic Susceptibility} 
\begin{figure}[bt]
\begin{center}
  \includegraphics{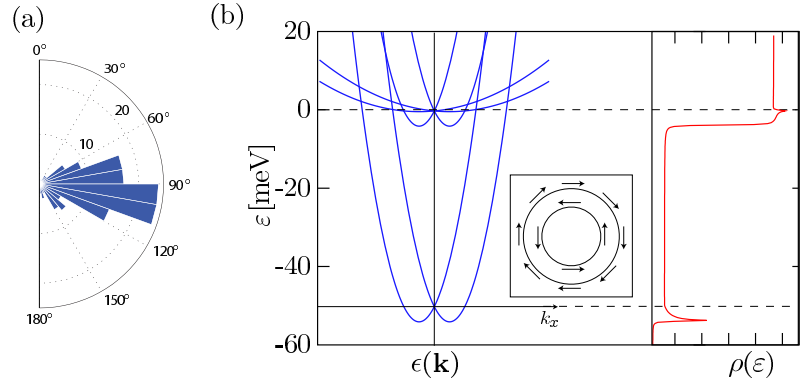}
\end{center}
\caption{(a) The distribution of the dipole-moment direction in terms of the angle from the $z$ axis as observed in reference \cite{bert:2011} (reproduced with permission). (b) Dispersion $\epsilon(\kk)$ and density of states $\rho(\varepsilon)$ of the three-band model equation \eref{eq:ham} along $k_x$. The two quasi-1D bands are shifted by $\Delta=50$meV compared to the 2D band. 
The inset shows a spin-texture along Rashba-spilt parabolic bands.}
\label{fig:disp}
\end{figure}

We use a three-band model for the Ti t$_{2g}$ orbitals in the $xy$ plane to describe the electronic structure of the interface\cite{popovic:2008}.  For now, we ignore the atomic spin-orbit coupling in order to gain more analytic insight. In the presence of an external magnetic field $\vec{H}$, the Hamiltonian reads
\begin{equation}
\HH = \HH_0 + \HH_{\rm soc}^{R} -\mu_B\vec{H}\cdot\vec{S}.
\label{eq:ham}
\end{equation}
Here, $\HH_0$ is the hopping Hamiltonian
\begin{equation}
\HH_0=\sum_{l,\kk,s} \xi_{l\kk}^{(0)} c_{l\kk s}^\dagger c^{\phantom{\dag}}_{l\kk s} 
\label{eq:xi}
\end{equation}
with bare dispersions $\xi_{l\kk}^{(0)} = k_x^2/2m_x^l + k_y^2/2m_y^l - \mu_l $ and $c^{\dag}_{l\kk s}$ creates an electron in band
$l=(1,2,3)\equiv$(d$_{xz}$, d$_{yz}$, d$_{xy}$) 
with momentum $\kk$, and spin $s$. We use mass parameters from reference \cite{santander-syro:2011}: the light masses $m=m_x^1=m_y^2=m_x^3=m_y^3 = 0.7 m_e$ and the heavy masses $M=m_y^1=m_x^2=15m_e$ with $m_e$ the electron mass. The chemical potentials are related by $\mu = \mu_1=\mu_2=\mu_3+\Delta$ and we use in the following $\Delta=50$meV. In the Zeeman term [the last term in equation \eref{eq:ham}], $\vec{S}=\sum_{l,\kk,s,s'}c^\dagger_{l\kk  s}\vec{\sigma}_{ss'}c^{\phantom{\dag}}_{l\kk s'}$ is the total spin with $\vec{\sigma}$ being the Pauli matrices. 
Finally, $\HH^R_{\rm soc}$ is the Rashba SOC at the interface due to the absence of the in-plane mirror symmetry, which
can phenomenologically be introduced as a relativistic effect due to an electric field $\vec{E}$ in $z$ direction:
The spin of an electron moving with  velocity $\vec{v}$ couples to an effective magnetic field
$ (\vec{E}/c \times \vec{v})$. Since the velocity ${\bf v}_l=\partial \xi^{(0)}_{l\kk}/\partial \kk$ for an electron in band $l$  is $\kk$-dependent, 
the Rashba SOC is 
\begin{equation}
\HH_{\rm soc}^{R} = \alpha\! \sum_{l,\kk,s, s'} \vec{g}_{l\kk}\cdot(c^{\dag}_{l\kk s}\vec{\sigma}_{ss'}c^{\phantom{\dag}}_{l\kk s'}), 
\label{eq:rashba}
\end{equation}
where $\vec{g}_{l\kk} = (v_{l,y}, -v_{l,x}, 0)$ and the overall scale $\alpha\approx 10^{-11}$eVm\cite{caviglia:2010}. The Rashba term  $\HH^R_{\rm soc}$ changes the zero-field bandstructure 
by splitting the spin degeneracy of the individual bands, as shown in 
Fig.~\ref{fig:disp}(b). 
Notice how different bands have different band-edge configurations: a ring of lowest energy momenta for the two-dimensional (\twod) $d_{xy}$ band and a saddle point for the two quasi-one-dimensional (\oned) bands stemming from the $d_{xz}$ and $d_{yz}$ orbitals. This difference is due to the isotropic (anisotropic) momentum dependence of the Rashba coupling $\vec{g}_{l\kk}$ for the \twod (\oned) bands and leads to different types of divergences in the density of states $\rho(\varepsilon)$ at the respective band edges: A $1/\sqrt{\varepsilon}$ divergence at the bottom of the $d_{xy}$ band and a logarithmic divergence near the bottom of the $d_{xz}$, $d_{yz}$ bands. 
In the regime of low electron densities, this system can thus have instabilities to broken-symmetry phases even for weak interactions.
 
Now we turn to the impact of $\HH^R_{\rm soc}$ on the in-field ($\vec{H}\neq0$) bandstructure, which leads to one of our main results: the anisotropy in the bare uniform susceptibility near band edges.
In order to calculate the bare susceptibility, we first diagonalize the Hamiltonian \eref{eq:ham} for each momentum $\kk$ to obtain the in-field spectrum $\xi_{\kk\nu}(\vec{H})$, where $\nu=1,\dots, 6$. Then, the (diagonal) susceptibility is given through
$\chi_{i} = \left.\frac{\partial^2\omega}{\partial H_i^2}\right|_{\vec{H}=0}$, with $\omega=\Omega/N$ the grand potential per lattice site,
\begin{equation}
\fl
\qquad\chi_{i} = \frac1N\sum_{\nu,\kk}\Big\{\frac{1}{4T\cosh[\xi_{\kk\nu}(\vec{H})/(2T)]^2}\Big[\frac{\partial \xi_{\kk\nu}(\vec{H})}{\partial H_i}\Big]^2
- n_{\rm F}[\xi_{\kk\nu}(\vec{H})]\Big[\frac{\partial^2\xi_{\kk\nu}(\vec{H})}{(\partial H_i)^2}\Big]\Big\}|_{\vec{H} =0},
\label{eq:chiii}
\end{equation}
where $T$ is the temperature and $n_{\rm F}(\xi)$ is the Fermi distribution function.
In the absence of the Rashba term, the Hamiltonian \eref{eq:ham} is diagonal for the spin-quantisation direction parallel to $\vec{H}$ and $\xi_{\kk\nu}(\vec{H})$ is linear in $\vec{H}$. Hence, only the first term in equation \eref{eq:chiii} contributes to the susceptibility: the usual Pauli susceptibility.
Since the Rashba term $\HH^R_{\rm soc}$ does not commute with the Zeeman term, $\xi_{\kk\nu}(\vec{H})$ is not linear in $\vec{H}$ once $\HH^R_{\rm soc}$ is present and the second term of equation \eref{eq:chiii}, the so-called van Vleck susceptibility, becomes non-zero. The direction-dependent balance between the two contributions determines a possible anisotropy $\chi_z\neq \chi_x$. 

As the Hamiltonian in equation \eref{eq:ham} is block diagonal in the orbital basis,
the total bare susceptibility is the sum of contributions $\chi_{i}(l)$ from each orbital $l$. 
We start with the contribution from the \oned orbitals. The two $d_{xz}$ bands have dispersions:
\begin{equation}
\xi^{xz}_{\kk \pm}(\vec{H}) = \frac{k_x^2}{2m} + \frac{k_y^2}{2M} - \mu \pm|(\alpha \vec{g}_{xz,\kk} - \mu_{\rm B}\vec{H})|.
\label{eq:xixz}
\end{equation}
They contribute to the total susceptibility through
\begin{equation}
\chi^{\rm P}_{i}(xz) = \mu_{\rm B}^2\sum_{\kk,\pm} (\hat{g}_{xz,\kk}^i)^2\frac{1}{4T\cosh[\xi^{xz}_{\kk\pm}(0)/(2T)]^2}
\label{eq:chip}
\end{equation}
and
\begin{equation}
\chi^{\rm vV}_{i}(xz) = \mu_{\rm B}^{2}\sum_{\kk}[1-(\hat{g}_{xz,\kk}^i)^2] \frac{n_{\rm F}[\xi^{xz}_{\kk-}(0)]-n_{\rm F}[\xi^{xz}_{\kk+}(0)]}{|\vec{g}_{xz,\kk}|}
\label{eq:chivV}
\end{equation}
with $\hat{g}_{xz,\kk}$ the unit vector along $\vec{g}_{xz,\kk}$. 
For $T\rightarrow 0$, we substitute $\tilde{k}_y = (M/m) k_y$ and change the sums in equations \eref{eq:chip} and \eref{eq:chivV} into (cylindrical) integrals.
For $\mu>0$, we obtain
\begin{equation}
\chi^{\rm P}_{i}(xz) =\frac{\mu_{\rm B}^2 M}{2\pi^2}\int d\phi
\frac{(\tilde{g}_{xz, \phi}^i)^2}{\cos^2\phi + \frac M m \sin^2\phi}
\label{eq:chipg0}
\end{equation}
and
\begin{equation}
\chi^{\rm vV}_{i}(xz) =\frac{\mu_{\rm B}^2 M}{2\pi^2} \int d\phi\frac{1-(\tilde{g}_{xz, \phi}^i)^2}{\cos^2\phi + \frac M m \sin^2\phi}.
\label{eq:chivVg0}
\end{equation}
with $\tilde{g}_{xz, \phi}^{i} = (\sin\phi, -\cos\phi, 0)$ and $\phi$ is the angle relative to the crystalline $x$-axis.
Clearly, the total  d$_{xz}$ contribution to the susceptibility $\chi_i(xz)= \mu_B^2\sqrt{mM}/\pi$ is independent of the field direction $i$ or the Rashba SOC strength $\alpha$. 

On the other hand, near the band edge of the one-dimensional bands, i.e., $-\alpha^2/2m<\mu<0$, the susceptibilities in equations \eref{eq:chip} and \eref{eq:chivV} yield
\begin{equation}
  \chi^{\rm P}_{i}(xz) =\frac{\mu_{\rm B}^2 M}{2\pi^2}\int d\phi
  \frac{(\tilde{g}_{xz, \phi}^i)^2}{\cos^2\phi + \frac M m \sin^2\phi}  [1 + 2\mu m(\cos^2\phi + \frac M m \sin^2\phi)/\alpha^2]^{-1/2}
  \label{eq:chipl0}
\end{equation}
and
\begin{equation}
  \chi^{\rm vV}_{i}(xz) =\frac{\mu_{\rm B}^2 M}{2\pi^2} \int d\phi\frac{1-(\tilde{g}_{xz, \phi}^i)^2}{\cos^2\phi + \frac M m \sin^2\phi}[1 + 2\mu m(\cos^2\phi + \frac M m \sin^2\phi)/\alpha^2]^{1/2}.
  \label{eq:chivVl0}
\end{equation}
The total contribution to the susceptibility is thus anisotropic at the band edge.
\begin{figure}[tb]
  \begin{center}
    \includegraphics{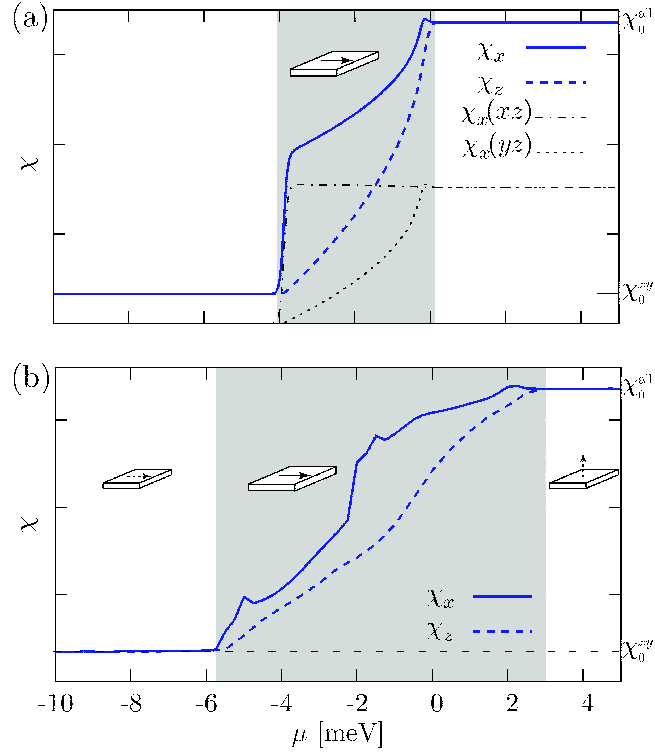}
  \end{center}
  \caption{In-plane (solid line) and out-of-plane (dashed line) total spin susceptibility for the three-band model (a) without and (b) with atomic SOC ($\alpha^{\rm at}=5$meV) with the gray area denoting the region with anisotropic susceptibility. The scale in these plots is given by $\chi_0^{xy} = \mu^2_{\rm B} m /\pi$ and $\chi_{0}^{\rm all}= \chi_{0}^{xy} + 2\mu_{\rm B}^2 \sqrt{Mm}/\pi$ (right vertical axes). In (a), the contributions of the two \oned bands for $\chi_x$ are also shown separately. 
  The arrows in the insets denote the preferred magnetisation direction, where in (b) also the anisotropy outside the 'Rashba' regime due to atomic SOC is shown.
  }
  \label{fig:susc}
\end{figure}

For the total bare susceptibility, we also need to consider the \twod orbital d$_{xy}$ contribution. We can read off $\chi_i(xy)$ from the \oned contribution $\chi_i(xz)$ discussed above by setting $M=m$ and shifting the band edge by $\Delta$. Therefore, $\chi_i(xy)=\chi_0^{xy}$ is isotropic for $\mu>-\Delta$. 
When contributions from all the orbitals are combined, the total susceptibility shows two regions in the chemical potential where $\chi_x=\chi_y>\chi_z$: near the band edge of the \twod bands and that of the \oned bands. Fig.~\ref{fig:susc}(a) shows the total susceptibility for the three bands near the \oned band edge as a function of chemical potential, where we shaded the anisotropic region.

Fig.~\ref{fig:susc}(b) summarises two ways atomic SOC impacts the phase diagram: 
(i) it widens the region of anisotropic susceptibility (grey region), (ii) it causes anisotropy in the spin direction by aligning the spin with the largest angular momentum when the susceptibility is isotropic (white region).
The first effect results from adding  
the atomic SOC, 
\begin{equation}
  \mathcal{H}_{\rm SOC}=i \frac{\alpha^{\rm at}}{2}\sum_{lmn}\epsilon_{lmn}\sum_{\kk, s, s'}c^{\dag}_{l\kk s}c^{\phantom{\dag}}_{m\kk s'}\sigma^n_{ss'}
  \label{eq:atomic}
\end{equation}
to the Hamiltonian \eref{eq:ham} and evaluating the susceptibility 
~\eref{eq:chiii} numerically. 
For the second effect, the atomic SOC is treated as a perturbation in the magnetic phase\cite{laan:1998}. 
This again predicts in-plane magnetisation below the grey region where only the \twod band is occupied, but a switch to out-of-plane magnetisation above this region.

In order to compare our phase diagram with experiments, we need to  translate the  chemical potential to a gate voltage.
While such a translation is non-trivial, Hall measurements under gate-voltage sweeps can offer hints as to where the as-grown samples lie. Joshua \textit{et al.} \cite{joshua:2011} showed that the interface acquires heavy carriers when under a gate voltage, which indicates that the as-grown samples are near the \oned band edge. 
Though the grey region is narrow in Fig.~\ref{fig:susc}(b), the density changes  
by a factor of $\approx 2$ in this range. This pushes the density upper bound for in-plane magnetisation substantially, consistent with experiments\cite{kalisky:2012}. 


\begin{figure}[tb]
  \begin{center}
    \includegraphics{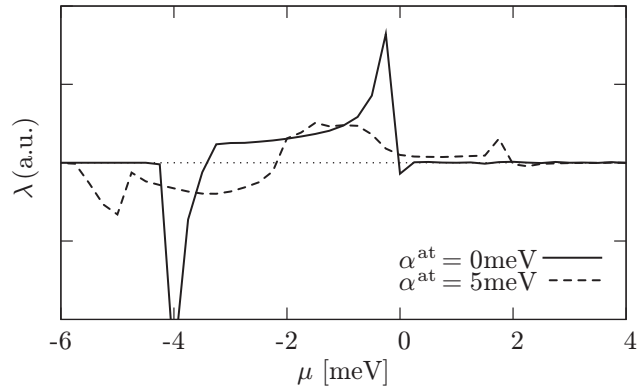}
  \end{center}
  \caption{Coupling constant $\lambda\propto\chi_{xx\eta}$ for coupling $\eta=n_{1}-n_{2}$ to $M_x^2 - M_y^2$ as a function of chemical potential for different atomic SOC parameters. Note that $\lambda$ for atomic spin-orbit coupling only would in this plot not significantly deviate from the zero line (thin dotted line).}
  \label{fig:fmeta}
\end{figure}
\section{Orbital Order and Magnetisation} 
In the regime of low electron densities, the van Hove singularity can promote broken symmetry phases in the presence of suitable interactions. While we focused so far on a magnetic instability, which could for instance be driven by repulsive intra-orbital interactions, we now turn to orbital-ordering possibilities. 
Even though the three-fold degeneracy of the $t_{2g}$ orbitals is already broken due to the interface symmetry, the $d_{xz}$ and $d_{yz}$ orbitals remain degenerate. A spontaneous orbital symmetry breaking described by a non-vanishing order parameter $\eta \equiv n_{1} - n_{2}$, with $n_{1/2}$ the occupation of the $d_{xz}/d_{yz}$ orbital, can be driven by inter-orbital repulsive interactions\cite{raghu:2009}.
We first analyse the tendency toward such an instability by studying the (bare) nematic susceptibility. For this purpose, we introduce the field $H_{\eta}$ conjugate to $\eta$, which enters the Hamiltonian \eref{eq:ham} through $\mu_{1/2}=\mu \pm H_{\eta}$, i.e. it acts through an opposite shift of the chemical potential for the two orbitals. The nematic susceptibility then yields
\begin{equation}
  \chi_{\eta} = \left.\frac{\partial^2\omega}{\partial H_{\eta}^2}\right|_{H_\eta = 0}=\frac{\partial^2 \omega}{\partial\mu_1^2} + \frac{\partial^2\omega}{\partial \mu_2^2}
  \label{eq:chieta}
\end{equation}
and is given by the contribution of the $d_{xz}$ and $d_{yz}$ orbitals to the density of states. Hence, the van Hove singularity near the band edge of the quasi-1D bands[see Fig.\ref{fig:disp}(b)] allows for a nematic order for sufficiently strong inter-orbital interaction\cite{raghu:2009}.

Next, we consider the coexistence of nematic and magnetic order.
While a system with $C_4$ symmetry does not allow for coupling of $\eta$ and $|\vec{M}|$ directly, there is an allowed tri-linear coupling between $\eta$ and an in-plane anisotropy $M_x^2 - M_y^2$, as both
acquire a factor of $-1$ under C$_4$ rotation. Specifically, a coupling of the form $\lambda(M_x^2 - M_y^2)\eta$ enters the free energy.
Given the Hamiltonian \eref{eq:ham}, we can explicitly calculate the coupling constant $\lambda$.
It is given by the generalised susceptibility 
\begin{equation}
  \lambda \propto \chi_{\eta xx} = \Big(\frac{\partial^3 \omega }{\partial H_\eta\partial H_x^2}\Big)_{|\vec{H}|=0},
  \label{eq:etasusc}
\end{equation}
which measures the change in the spin susceptibility $\chi_{x}$ upon shifting the chemical potential of the two \oned bands against each other. 
Figure~\ref{fig:fmeta} shows the result with and without atomic SOC in the presence of Rashba SOC. (We find the coupling to be negligible without Rashba SOC.) The coupling becomes substantial in the above-identified density range with the in-plane preference in the bare spin susceptibility. 
This is due to the 
inequivalence between the two \oned band contributions to $\chi_x$ shown as
dotted and dash/dotted lines in Fig.~\ref{fig:susc}(a), more  
specifically the difference in the slope of $\chi_x(xz)$ and $\chi_x(yz)$.  
The sign of $\lambda$ determines the relative sign between  $\eta$ and $(M_x^2 - M_y^2)$ and whether the majority orbitals will be along or perpendicular to the magnetisation axis. 
 Notice how our result shown in Fig.~\ref{fig:fmeta} predicts a change of the sign of $\lambda$ over the chemical potential range exhibiting the magnetisation anisotropy. 
 
Given the observed magnetism, we finally investigate the effects of coupling between magnetic order and nematic fluctuation associated with nearby nematic phase.
In particular, we may ask on the one hand how the proximity to a nematic instability influences the in-plane magnetisation, and on the other hand, how a magnetisation effects the orbital ordering.
Assuming an XY ferromagnet in the absence of a coupling to nematic order, the Landau free energy becomes
\begin{equation}
  f(M_x, M_y, \eta) = f_0 + \frac{a(T)}2 |\vec{M}|^2 + \frac b4 |\vec{M}|^4 + \lambda(M_x^2 - M_y^2)\eta  + \frac{a_{\eta}}2 \eta^2+\cdots,
  \label{eq:f0}
\end{equation}
where $M_{x}$ ($M_y$) is the magnetisation along the crystalline $x$ ($y$) axis. 
Since we assume no independent instability towards an orbital-ordered nematic, $a_{\eta}>0$\footnote{Even if strong correlation drives orbital order there will
  be no qualitative change in the effect of the coupling $\lambda $ in driving
  the in-plane magnetisation anisotropy.}.
Integrating out $\eta$ by minimising the free energy with 
\begin{equation}
  \eta = -\frac{\lambda}{a_{\eta}}(M_x^2 - M_y^2),
  \label{eq:eta}
\end{equation}
the free energy for the magnetisation becomes
\begin{equation}
  f(M_x, M_y) = f_0 + \frac{a(T)}2 |\vec{M}|^2 + \frac b4|\vec{M}|^4 -\frac{\lambda^2}{a_{\eta}}(M_x^2-M_y^2)^2.  
  \label{eq:f1}
\end{equation}
For finite $\lambda$,  the coupling to orbital order therefore locks the  
magnetisation along one of the crystal axes, leading to an additional in-plane anisotropy.

For the orbital ordering, the magnetisation anisotropy acts as a driving field. For $a_{\eta}$ very small, i.e., the system close to an instability, we should include additional terms to the free energy for $\eta$,
\begin{equation}
  f(\eta;\vec{M}) = \frac{a_{\eta}}2\eta^2 + \frac{b_{\eta}}4\eta^4 + \frac{c_{\eta}}6 \eta^6 +\lambda(M^2_x - M_y^2)\eta+\cdots.
  \label{eq:freeeta}
\end{equation}
For $b_{\eta}<0$, the system undergoes either a metanematic crossover, or a first-order transition at a critical magnetisation. This is in analogy to metamagnetic transitions observed in systems close to ferromagnetism\cite{levitin:88} and could here indirectly be driven by an applied magnetic field.

\section{Concluding Remarks} 
We have shown that the combination of Rashba SOC and atomic SOC leads to an electron-density dependent magnetisation anisotropy in LAO/STO interfaces. While experiments so far appear to lie in the in-plane-magnetisation region, we predict a switch to out-of-plane magnetisation at sufficiently high gate voltages. 
We have identified a regime near the band edge with anisotropic susceptibility
as a non-trivial effect of the Rashba SOC in a low carrier density system. The high density of states near the band edge in principle also allows for a spontaneous orbital ordering
and we predict in this regime an enhanced coupling between the magnetisation direction and this kind of nematic order. 
This coupling locks the in-plane magnetisation direction to be along one of the crystal axes and promotes Ising nematicity. 


Next, we comment on the issue of heterogeneity detected in reference \cite{bert:2011}. The observed heterogeneity is likely driven by both extrinsic and intrinsic effects. For instance, a recent study showed that strong Rashba SOC can promote phase separation\cite{caprara:2012}. The proposed magnetisation-nematicity coupling has important consequences in both extrinsic and intrinsic fronts. On the one hand, oxygen vacancies and other spatial inhomogeneities act as a random field for the Ising nematic and in turn cause a distribution of moment directions. On the other hand, the reduction of the magnetic order parameter symmetry due to the coupling changes the type of magnetic textures and their energetics. Furthermore, we expect the coupling to cause non-trivial in-plane anisotropy in the magneto transport \footnote{Fischer et al., in preparation}. Moreover, the sign change in the coupling $\lambda$ could be observed through the rotation in the dominant direction upon gate voltage sweep.
Finally, we note that the range in the density with magnetisation-nematicity near the band edge of the quasi-one-dimensional bands has also been shown to exhibit critical scaling in recent Hall measurement\cite{joshua:2011}.

\ack
We are grateful to J.~Bert, H.~Hwang, B.~Kalisky, and K.~Moler for useful discussions. 
MHF and E-AK acknowledge support from NSF Grant DMR-0955822 and from NSF Grant DMR-1120296 to the Cornell Center for Materials Research. SR acknowledges support from the LDRD program at SLAC and the Alfred. P. Sloan Research fellowship.


\section*{References}
\begin{thebibliography}{20}

\bibitem{zutic:2004}
Igor \ifmmode \check{Z}\else \v{Z}\fi{}uti\ifmmode~\acute{c}\else \'{c}\fi{},
  Jaroslav Fabian, and S.~Das~Sarma.
\newblock Spintronics: Fundamentals and applications.
\newblock {\em Rev. Mod. Phys.}, 76:323--410, Apr 2004.

\bibitem{winkler:2003}
R~Winkler.
\newblock {\em Spin-orbit coupling effects in two-dimensional electron and hole
  systems}, volume 191.
\newblock Springer-Verlag Berlin, 2003.

\bibitem{berg:2012}
Erez Berg, Mark~S. Rudner, and Steven~A. Kivelson.
\newblock Electronic liquid crystalline phases in a spin-orbit coupled
  two-dimensional electron gas.
\newblock {\em Phys. Rev. B}, 85:035116, Jan 2012.

\bibitem{hwang:2012}
H.~Y. Hwang, Y.~Iwasa, M.~Kawasaki, B.~Keimer, N.~Nagaosa, and Y.~Tokura.
\newblock Emergent phenomena at oxide interfaces.
\newblock {\em Nat Mater}, 11(2):103--113, 02 2012.

\bibitem{brinkman:2007}
A.~Brinkman, M.~Huijben, M.~van Zalk, J.~Huijben, U.~Zeitler, J.~C. Maan, W.~G.
  van~der Wiel, G.~Rijnders, D.~H.~A. Blank, and H.~Hilgenkamp.
\newblock Magnetic effects at the interface between non-magnetic oxides.
\newblock {\em Nat Mater}, 6(7):493--496, 07 2007.

\bibitem{caviglia:2008}
A.~D. Caviglia, S.~Gariglio, N.~Reyren, D.~Jaccard, T.~Schneider, M.~Gabay,
  S.~Thiel, G.~Hammerl, J.~Mannhart, and J.~M. Triscone.
\newblock Electric field control of the LaAlO$_3$/SrTiO$_3$ interface ground state.
\newblock {\em Nature}, 456(7222):624--627, 12 2008.

\bibitem{reyren:2007}
N.~Reyren, S.~Thiel, A.~D. Caviglia, L.~Fitting Kourkoutis, G.~Hammerl,
  C.~Richter, C.~W. Schneider, T.~Kopp, A.-S. R√ºetschi, D.~Jaccard,
  M.~Gabay, D.~A. Muller, J.-M. Triscone, and J.~Mannhart.
\newblock Superconducting interfaces between insulating oxides.
\newblock {\em Science}, 317(5842):1196--1199, 2007.

\bibitem{bert:2011}
Julie~A. Bert, Beena Kalisky, Christopher Bell, Minu Kim, Yasuyuki Hikita,
  Harold~Y. Hwang, and Kathryn~A. Moler.
\newblock Direct imaging of the coexistence of ferromagnetism and
  superconductivity at the LaAlO$_3$/SrTiO$_3$ interface.
\newblock {\em Nat Phys}, 7(10):767--771, 10 2011.

\bibitem{li:2011}
Lu~Li, C.~Richter, J.~Mannhart, and R.~C. Ashoori.
\newblock Coexistence of magnetic order and two-dimensional superconductivity
  at LaAlO$_3$/SrTiO$_3$ interfaces.
\newblock {\em Nat Phys}, 7(10):762--766, 10 2011.

\bibitem{bland:2005}
J~A~C Bland and Bretislav Heinrich.
\newblock {\em Ultrathin magnetic structure I}.
\newblock Springer Berlin / Heidelberg, 2005.

\bibitem{draaisma:1988}
H.~J.~G. Draaisma and W.~J.~M. de~Jonge.
\newblock Surface and volume anisotropy from dipole-dipole interactions in
  ultrathin ferromagnetic films.
\newblock {\em Journal of Applied Physics}, 64(7):3610--3613, 1988.

\bibitem{laan:1998}
Gerrit van~der Laan.
\newblock Microscopic origin of magnetocrystalline anisotropy in transition
  metal thin films.
\newblock {\em Journal of Physics: Condensed Matter}, 10(14):3239, 1998.

\bibitem{joshua:2011}
A.~{Joshua}, S.~{Pecker}, J.~{Ruhman}, E.~{Altman}, and S.~{Ilani}.
\newblock {A Universal Critical Density Underlying the Physics of Electrons at
  the LaAlO$_3$/SrTiO$_3$ Interface}.
\newblock {\em Nature Commun.}, 3:1129, 2012

\bibitem{popovic:2008}
Zoran~S. Popovi\ifmmode~\acute{c}\else \'{c}\fi{}, Sashi Satpathy, and
  Richard~M. Martin.
\newblock Origin of the two-dimensional electron gas carrier density at the
  LaAlO$_3$ on SrTiO$_3$ interface.
\newblock {\em Phys. Rev. Lett.}, 101(25):256801, Dec 2008.

\bibitem{santander-syro:2011}
A.~F. Santander-Syro, O.~Copie, T.~Kondo, F.~Fortuna, S.~Pailhes, R.~Weht,
  X.~G. Qiu, F.~Bertran, A.~Nicolaou, A.~Taleb-Ibrahimi, P.~Le~Fevre,
  G.~Herranz, M.~Bibes, N.~Reyren, Y.~Apertet, P.~Lecoeur, A.~Barthelemy, and
  M.~J. Rozenberg.
\newblock Two-dimensional electron gas with universal subbands at the surface
  of SrTiO$_3$.
\newblock {\em Nature}, 469(7329):189--193, 01 2011.

\bibitem{caviglia:2010}
A.~D. Caviglia, M.~Gabay, S.~Gariglio, N.~Reyren, C.~Cancellieri, and J.-M.
  Triscone.
\newblock Tunable Rashba spin-orbit interaction at oxide interfaces.
\newblock {\em Phys. Rev. Lett.}, 104:126803, Mar 2010.

\bibitem{kalisky:2012}
Beena Kalisky, Julie~A. Bert, Brannon~B. Klopfer, Christopher Bell, Hiroki~K.
  Sato, Masayuki Hosoda, Yasuyuki Hikita, Harold~Y. Hwang, and Kathryn~A.
  Moler.
\newblock Critical thickness for ferromagnetism in LaAlO$_3$/SrTiO$_3$
  heterostructures.
\newblock {\em Nat Commun}, 3:922, 06 2012.

\bibitem{raghu:2009}
S.~Raghu, A.~Paramekanti, E~A. Kim, R.~A. Borzi, S.~A. Grigera, A.~P.
  Mackenzie, and S.~A. Kivelson.
\newblock Microscopic theory of the nematic phase in Sr$_3$Ru$_2$O$_7$.
\newblock {\em Phys. Rev. B}, 79(21):214402, 2009.

\bibitem{levitin:88}
R~Z Levitin and A~S Markosyan.
\newblock Itinerant metamagnetism.
\newblock {\em Soviet Physics Uspekhi}, 31(8):730, 1988.

\bibitem{caprara:2012}
S.~Caprara, F.~Peronaci, and M.~Grilli.
\newblock Intrinsic instability of electronic interfaces with strong Rashba
  coupling.
\newblock {\em Phys. Rev. Lett.}, 109:196401, Nov 2012.

\end{thebibliography}
\end{document}